\documentstyle[11pt]{article}
\newcommand{\non}{\nonumber}

\newcommand{\la}{\langle}
\newcommand{\ra}{\rangle}

\newcommand{\pa}{\partial}
\renewcommand{\a}{\alpha}
\renewcommand{\b}{\beta}
\renewcommand{\c}{\gamma}

\newcommand{\m}{\mu}
\newcommand{\n}{\nu}
\newcommand{\bpsi}{{\bar{\psi}}}

\begin{document}

\rightline{UCY-PHY 95/11}
\vspace{1cm}
\begin{center}
\large
{\bf{Operator Product Expansions and Consistency Relations in a $O(N)$  
Invariant Fermionic CFT for $2<d<4$}}
\vspace{2cm}
\normalsize
\\
{\bf{Anastasios C. Petkou}}\footnote{e-mail: petkou@pauli.ns.ucy.ac.cy} \\ 
Department of Natural Sciences \\ University of Cyprus\\ P.O. Box 537
Nicosia \\ Cyprus
\end{center}

\vspace{2cm}
\begin{abstract}
A conformally invariant theory of Majorana fermions in $2<d<4$ with $O(N)$
symmetry  is  studied using  Operator  Product Expansions  and
consistency relations based on  the cancellation of shadow
singularities. The critical coupling $G_{*}$ of the theory is
calculated to leading order in $1/N$. This  value is then used to
reproduce the $O(1/N)$ correction for the anomalous dimension of the
fermion field as evidence for the validity of our approach to
conformal field theory in $d>2$. 
\end{abstract}
\newpage

\subsection*{Introduction}

Conformal field theory (CFT) methods have been proved powerful in
describing   the  
critical properties of systems in dimensions $d>2$
\cite{Vasiliev,Gracey,Semenoff,Ruhl}.  Recently, \cite{tasos1,tasos2}, we
proposed a systematic treatment of the conformally invariant $O(N)$
vector model in $2<d<4$ based on Operator Product Expansions (OPE's) and
consistency relations requiring the cancellation of shadow
singularities. In \cite{tasos1,tasos2} we presented strong evidence that
the proposed treatment gives the same results with the customary
$1/N$ expansion techniques for the anomalous
dimensions of the fields in the $O(N)$ vector model, (see \cite{Ruhl}
and  references therein),
while it proves more suitable for the evaluation of the important quantities
$C_{T}$ and $C_{J}$. It also unveils an interesting  duality property of
the model.

Our approach to  the $O(N)$ vector model in \cite{tasos1,tasos2} can be
viewed as an effort to obtain  dynamical information for a   CFT in 
$d>2$ by algebraic methods, i.e.
OPE's and consistency relations, without using a Lagrangian. If
correct, our  
approach  must be generally
applicable to all CFT's in $d>2$.  In this letter we apply our
algebraic approach to a  fermionic $O(N)$   
invariant  model   in $2<d<4$ and we show that it gives the same
result with $1/N$ expansion techniques for the anomalous dimension of
the fermion field. The calculational details will be
presented in a forthcoming publication \cite{tasos3}.

\subsection*{OPE's and Consistency Relations in a
Fermionic $O(N)$ Invariant CFT in $2<d<4$}

We consider a Euclidean fermionic CFT having as fundamental
fields the
$N$-component Majorana fermions       
$\psi^{\a}(x)$, ${\bar{\psi}}^{\a}(x)$, $\a=1,2,..,N$. The Majorana
condition $\psi(x)=C\,\bar{\psi}^{T}(x)$ ensures that the fermions are
effectively real and the internal symmetry group at the fixed point
maximal, the latter is taken here to be $O(N)$. We use Hermitean gamma
matrices  
${\gamma_{\mu}\gamma_{\nu}+\gamma_{\nu}\gamma_{\nu}=2\,\delta_{\mu\nu}\,{
\bf{1}}}$. The charge conjugation matrix $C$  is
defined  \footnote{See e.g. \cite{Zinn-Justin,DeWit}. $C$ has these
properties for $d=2,3,4$ when the dimension of the gamma matrices is
respectively $2,2,4$. For general $d$ we assume $C^{T}=-C$ which
is always  possible if the representation of the gamma matrices is
reducible.} through
$C^{T}=-C^{-1}$, $C\,\gamma_{\mu}^{T}\,C^{-1}=-\gamma_{\mu}$. 
The  conformally invariant
two-point function $\la\psi^{\a}(x_{1}){\bar{\psi}}^{\b}(x_{2})\ra$ takes
the form \cite{Gracey,Semenoff}
\begin{equation}
  \la\psi^{\a}(x_{1}){\bar{\psi}}^{\b}(x_{2})\ra
=C_{\psi}\,\frac{\not\!x_{12}}{(x_{12}^{2})^{a}}\delta^{\a\b}
,\,\,\,\,\,\,
\,\,  \not\!x_{12}=\gamma\!\cdot\!x_{1}-\gamma\!\cdot\!x_{2}\,,\,\,\,\, a
= \eta+{\textstyle{\frac{1}{2}}}. 
\label{eq1}
\end{equation}
The normalisation constant $C_{\psi}$
can  be arbitrarily chosen and we henceforth set it to
one \footnote{Generally speaking, in CFT the relative
normalisation of two- and three-point functions depends on the
dynamics. Our approach henceforth 
corresponds to fixing  to unity the normalisation of the two-point functions
and taking  the normalisation of the three-point functions (couplings)
to be the dynamical variables. Note however that the normalisation of the
two-point functions of conserved currents, such as  the energy momentun
tensor $T_{\m\n}(x)$,  is not arbitrary but it is
fixed by the Ward identities of the theory \cite{tasos1,tasos2,Cardy}.}. 

Discussions of CFT's usually concentrate on $n$-point
functions  of
quasiprimary fields  with $n\geq 4$ since the functional form of the
two- and
three-point functions is fixed up to some arbitrary constants
\cite{tasos1}. In this letter we consider the
four-point  function 
\begin{eqnarray}
\hspace{-1.5cm} \la\psi^{\a}_{i}(x_{1})\!\!\!\!\!\!\!&
&\!\!\!\!\!\!\!\bpsi^{\b}_{j}(x_{2})\psi^{\gamma}_{k}(x_{3})
 \bpsi^{\delta}_{l}(x_{4})\ra  \,\,\equiv\,\, 
\Psi^{\a\b\c\delta}_{ijkl}(x_{1},x_{2},x_{3},x_{4}) \nonumber \\
 & = &\Psi_{ij,kl}(x_{1},x_{2};x_{3},x_{4})\,\delta^{\a\b}\delta^{c\delta} 
+\Psi_{in,ml}(x_{1};x_{3},x_{2};x_{4})\,C_{kn}\,C^{-1}_{mj}\,\delta^{\a\c}
\,\delta^{\b\delta}
\nonumber \\
 & & \,-\,\Psi_{il,kj}(x_{1},x_{4};x_{3},x_{2})
\,\delta^{\a\delta} \delta^{\c\b}\,,
\label{eq2}
\end{eqnarray}
where $i,j,k,l,n,m$ are spinor indices. We
may cast (\ref{eq2}) in a more convenient for what follows form as
\begin{eqnarray}
\Psi_{ijkl}^{\a\b\c\delta}(x_{1},x_{2},x_{3},x_{4}) & = &
F^{S}_{ijkl}(x_{1},..,x_{4})\,\delta^{\a\b}\delta^{\c\delta}+F^{V}_{ijkl}
(x_{1},..,x_{4})\,{\textstyle{\frac{1}{2}}}(\delta^{\a\c}\delta^{\b
\delta}
-\delta^{a\delta}\delta^{\b\c}) \nonumber  \\  
& &
{}+F^{T}_{ijkl}(x_{1},..,x_{4})\,{\textstyle{\frac{1}{2}}}(\delta^{ 
\a\c}\delta^{\b\delta}+\delta^{\a\delta}
\delta^{\b\c}-{\textstyle{\frac{2}{N}}}\delta^{\a\b}\delta^{\c\delta})\,,
\label{eq3}\\
F^{S}_{ijkl}(x_{1},..,x_{4}) & = &
\Psi_{ij,kl}(x_{1};x_{2},x_{3};x_{4})+{\textstyle{\frac{1}{N}}}
\Bigl[\Psi_{in,mj}
(x_{1};x_{3},x_{2};x_{4})\,C_{kn}\,C^{-1}_{mj} \non \\
 & & {}-\Psi_{il,kj}(x_{1};x_{4},x_{3};x_{2})\Bigl]\,,\label{eq4}\\
F^{V}_{ijkl}(x_{1},..,x_{4}) & = & \Psi_{in,mj}
(x_{1};x_{3},x_{2};x_{4})\,C_{kn}\,C^{-1}_{mj}+\Psi_{il,kj}(x_{1};x_{4},
x_{3};x_{2})\,,\label{eq5}\\
F^{T}_{ijkl}(x_{1},..,x_{4}) & = & \Psi_{in,mj}
(x_{1};x_{3},x_{2};x_{4})\,C_{kn}\,C^{-1}_{mj}-\Psi_{il,kj}(x_{1};x_{4},
x_{3};x_{2})\,,\label{eq6}
\end{eqnarray}

In correspondence with our treatment of the $O(N)$ vector model,
\cite{tasos1,tasos2},  we suggest
that we may evaluate (\ref{eq3}) on inserting into  it the  OPE of 
$\psi$  with $\bpsi$. The (infinite) quasiprimary fields   
appearing in such an  OPE  are
in principle unknown. However, as in
\cite{tasos1,tasos2}, we assume that this OPE is
qualitatively similar to the free field theory one, at least as far as
the most singular terms in the short distance limit are concerned.
In practice this means that we can write down an ansatz for the
leading terms in  the OPE 
$\psi_{i}^{\a}(x_{1})\bpsi_{j}^{\b}(x_{2})$ based on a Taylor
expansion for $x_{12}^{2}\rightarrow 0$ as in free field
theory. It is crucial for our approach that a ``low-lying scalar
field''  $O(x)$, which is  $O(N)$ singlet with 
dimension $\eta_{o}$, $0<\eta_{o}<d$, appears in the general OPE
ansatz  as it
appears  in free field theory.  Substituting then our  ansatz into 
(\ref{eq3}) we obtain a short distance expansion for the
four-point function. Such an  expansion generally depends on a number
of parameters, 
namely the couplings and the dimensions of the fields appearing in the OPE
ansatz. 

Next, we construct  graphically the amplitude for the four-point
function (\ref{eq3}) in terms of skeleton graphs, having no self-energy
or vertex insertions, with internal lines corresponding to the full
two-point functions   $\la\psi\bpsi\ra$ and another one, that of a
scalar $O(N)$ 
singlet field $\tilde{O}(x)$ with dimension $\tilde{\eta_{o}}$,
$0<\tilde{\eta_{o}}<d$. The latter field is related
\cite{tasos1,tasos2} to the ``low-lying scalar field'' appearing in
our OPE ansatz. Symmetry factors are determined as in the
usual Feynman perturbation expansion. The triple vertices connecting
the lines in such graphs \footnote{The skeleton graph
expansion for the free theory corresponds simply to the disconnected
graphs e.g.  see \cite{tasos1}.} are the fully amputated  three-point
functions $\la\psi\bpsi\tilde{O}\ra$,  whose functional form being
completely determined from conformal invariance depend only  on the field
dimensions and a coupling $G_{*}$. The important  point of our
graphical construction is that such graphs represent conformally
invariant amplitudes as long  as the dimensions of the fields involved
in their construction satisfy a certain ``uniqueness'' condition
\cite{Mack}. Then, these graphs can in principle be
calculated  using 
the results in \cite{DEPP}. For the simplest
cases  we have found closed  analytical expressions
\cite{tasos1,tasos2,tasos3}.  Consistency then of
the graphical and the OPE evaluation of the four-point function
(\ref{eq3}) determines the couplings  and the field
dimensions in the theory in the context of a self-consistent $1/N$ expansion.
 
Having outlined the general idea behind our approach to CFT in $d>2$
we concentrate on our fermionic theory. For general $d$ an OPE ansatz
consistent 
with $O(N)$ and conformal 
symmetry can be   taken as 
\begin{eqnarray}
\psi_{i}^{\a}(x_{1})\bpsi_{j}^{\b}(x_{2})
  & = &  \la\psi_{i}^{\a}(x_{1})\bpsi_{j}^{\b}(x_{2})\ra \non \\
& + &
\sum_{O,J,S}\sum_{n}(\gamma_{[\mu_{1},..,\mu_{n}]})_{ij}
\Bigl[C^{k}_{[\mu_{1},.., 
\mu_{n}]}(x_{12},\pa_{2})\,O_{k}(x_{2})\,\delta
^{\a\b}\non \\
 & + & \,D^{k}_{[\mu_{1},..,
\mu_{n}]}(x_{12},\pa_{2})\,J^{\a\b}_{k}(x_{2})
+  E^{k}_{[\mu_{1},..,
\mu_{n}]}(x_{12},\pa_{2})\,S^{\a\b}_{k}(x_{2})\Bigl]\,,
\label{eq7}
\end{eqnarray}
where $[\,]$ denotes antisymmetrisation in the
corresponding indices and $k$ is a general spin label. The
$J^{\a\b}_{k}(x)$ fields are antisymmetric and the 
$S^{\a\b}_{k}(x)$ are symmetric and traceless in the $O(N)$ indices. Then,
from (\ref{eq3}) and (\ref{eq7}) it follows that $F^{S}_{ijkl}$,
$F^{V}_{ijkl}$ and  $F^{T}_{ijkl}$ receive separately 
contributions from the $O(x)_{k}$, $J^{\a\b}_{k}(x)$ and
$S^{\a\b}(x)_{k}$  fields.  

Basically, we  have expanded the
 left hand side of (\ref{eq7}) in the  complete basis
$\gamma_{[\mu_{1},..,\mu_{n}]}$ of the antisymmetrised products of the
gamma  matrices i.e.
${\bf{1}}_{ij}$, $(\gamma_{\mu})_{ij}$, $(\gamma_{[\mu_{1},\mu_{2}]})_{ij}
= \frac{1}{2}[\gamma_{\mu_{1}},\gamma_{\mu_{2}}]_{ij}$, e.t.c.  \cite{DeWit}.
For general $d$ this basis is infinite but it truncates when
$d=2,3,4$. The dimension of the gamma matrices is  $2^{d/2}\times
2^{d/2}$, ($2^{(d-1)/2}\times 2^{(d-1)/2}$ for odd $d$), when 
$d=2,3,4$ but it is essentially arbitrary for non-integer $d$ and in
this case   an
infinite number of antisymmetrised 
products may be present.

A special role in the OPE (\ref{eq7}) is played by the scalar
 $O(N)$ singlet field $O_{0}(x)\equiv O(x)$ with dimension $\eta_{o}$,
$0<\eta_{o}<d$. Actually, a basic assumption concerning the form of the
OPE (\ref{eq7}) is to require the existence of {\it{only one}} such a field
i.e. a ``low-lying scalar field'' giving more singular contributions than
the energy momentum tensor \footnote{Such a form for the OPE of CFT's
in four dimensions has been recently discussed in \cite{Grisaru}.}  
\cite{tasos1,tasos2}. The crucial 
point is then that from the conformally invariant form of the two- and
three-point functions 
\cite{Semenoff,tasos1,tasos2}
\begin{eqnarray}
\la O(x_{1})O(x_{2})\ra & = &
C_{O}\frac{1}{x_{12}^{2\eta_{o}}}\,,\label{eq8} \\
\la\psi^{\a}_{i}(x_{1})\bpsi^{\b}_{j}(x_{2})O(x_{3})\ra & = &
g_{\psi\bpsi
O}\frac{(\not\!x_{13}\!\not\!x_{23})_{ij}}{(x_{12}^{2})^{\eta-\frac{1}{2}\eta
_{o}}
(x_{13}^{2}x_{23}^{2})^{\frac{1}{2}\eta_{o}+\frac{1}{2}}}\delta^{\a\b}
\,\equiv \,
C_{ij}^{0}(x_{12},\pa_{2})\,\frac{C_{O}}{x_{23}^{2\eta_{o}}}\,\delta^{\a\b}\,,
\label{eq9} 
\end{eqnarray}
with $g_{\psi\bpsi O}$ the coupling, one can evaluate the OPE coefficients
$C^{0}_{ij}(x_{12},\pa_{2})$,  i.e. following   techniques
presented in detail in   
\cite{tasos1}, as 
\begin{eqnarray} 
C^{0}_{ij}(x_{12},\pa_{2}) &=& 
\frac{g_{\psi\bpsi
O}/C_{O}}{B(\frac{1}{2}\eta_{o}+\frac{1}{2},\frac{1}{2}\eta_{o}
-\frac{1}{2})}\frac{1}{(x_{12}^{2})^{a-\frac{1}{2}\eta_{o}-\frac{1}{2}}}\,
\non \\
 & & \hspace{-2cm}\times({\bf{1}}-{
\textstyle{\frac{1}{\eta_{o}-1} }}\not\!x_{12}\not\!\pa_{2})_{ij}
\int_{0}^{1}dt\,
t^{\frac{1}{2}\eta_{o}-\frac{1}{2}}(1-t)^{\frac{1}{2}\eta_{o}
-\frac{3}{2}}
\sum_{m=0}^{\infty}\frac{1}{m!}\frac{[-\frac{1}{4}
t(1-t)x_{12}^{2}]^{m}}
{(\eta_{o}+1-\mu)_{m}}\pa_{2}^{2m}e^{tx_{12}\!\cdot\!\pa_{2}}
\,,\label{eq10} 
\end{eqnarray}
where $(a)_{m} = \Gamma(\mu+a)/\Gamma(a)$, 
$\mu = d/2$ and the derivatives on the r.h.s. of (\ref{eq10}) are
taken with constant $|x_{12}|$. Henceforth we set  $C_{O}=1$. Note
that (\ref{eq10}) is of order
$O(x_{12}^{\eta_{o}-2\eta} )$, that is it includes the most singular
contribution of $O(x)$ in (\ref{eq7}) as $x_{12}^{2}\rightarrow 0$.

Substituting the OPE (\ref{eq7}) into the four-point function
(\ref{eq3}) we can in principle find expressions for $F_{ijkl}^{S}$,
$F^{V}_{ijkl}$ and $F^{T}_{ijkl}$. Using (\ref{eq10}) we obtain for
example
\begin{eqnarray}
F^{S}_{ijkl}(x_{1},..,x_{4}) & = &
\frac{(\not\!x_{12})_{ij}\otimes(\not\!x_{34})_{kl}}{ 
(x_{12}^{2}x_{34}^{2})^{a}}+g^{2}_{\psi\bpsi O}
\frac{x_{24}^{2}}{(x_{12}^{2}x_{34}^{2})^{a}}{\cal{A}}^{(\eta_{o})}_{ijkl}
(x_{1},..,x_{4})
+\cdots\,,\label{eq11}\\
{\cal{A}}^{(\eta_{o})}_{ijkl}(x_{1},..,x_{4}) & = &
({\bf{1}}-{\textstyle{\frac{1}{\eta_{o}-1}}}
\not\!x_{12}\!\not\!\partial_{2})_{ij}\otimes
({\bf{1}}-{\textstyle{
\frac{1}{\eta_{o}-1}}}
\not\!x_{34}\!\not\!\partial_{4})_{kl}\left[{\cal{H}}^{(\eta_{o})}(u,v)\right]
\,,\label{eq12}\\
{\cal{H}}_{F}^{(\eta_{o})}(u,v) & = &
v^{\frac{1}{2}\eta_{o}+\frac{1}{2}}\sum_{n=0}^{\infty}
\frac{v^{n}}{n!}\frac{(\frac{1}{2}\eta_{o}+\frac{1}{2})_{n}^{2}(\frac{1}{2}
\eta_{o}-\frac{1}{2})_{n}^{2}}
{(\eta_{o}+1-\mu)_{n}(\eta_{o})_{2n}} \non \\
 & & \times{}_{2}F_{1}({\textstyle{\frac{1}{2}\eta_{o}+\frac{1}{2}}}
+n, {\textstyle{\frac{1}{2}\eta_{o}+\frac{1}{2}}} +n;
\eta_{o}+2n;1-{\textstyle{\frac{v}{u}}})\,, \label{eq13}
\end{eqnarray}
where we have  used the usual invariant ratios
$u=(x_{12}^{2}x_{34}^{2}/x_{13}^{2}x_{24}^{2})$,
$v=(x_{12}^{2}x_{34}^{2}/x_{14}^{2}x_{23}^{2})$.

The second term on the r.h.s of (\ref{eq11}) is the full contribution
of the $O(x)$ field to the four-point function. Here we shall only use
the most singular term of 
this contribution in the limit as $x_{12}^{2}$, $x_{34}^{2}\rightarrow
0$, or equivalently as $u$, $v\rightarrow 0$, namely
\begin{equation}
{\cal{A}}^{(\eta_{o})}_{ijkl}(x_{1},x_{2},x_{3},x_{4})\,\,\leadsto\,\,
{\bf{1}}_{ij}\otimes
{\bf{1}}_{kl}\,v^{\frac{1}{2}\eta_{o}+\frac{1}{2}}+
v^{\frac{1}{2}\eta_{o}+ \frac{1}{2}}\,{\cal{T}}_{ijkl}\,O(|x_{12}|,|x_{34}|)
\,.\label{eq14}
\end{equation}  

Another
important field  on the r.h.s of (\ref{eq7}) is the $O(N)$
conserved current appearing for example as\footnote{This term includes
the most singular contribution of $J_{\m}^{\a\b}(x)$ and it is all we
need for our subsequent calculations.}
$D^{1}(x_{12},\pa_{2})\,J^{\a\b}_{\mu}(x_{2})$ with dimension $d-1$
and spin 1. Its
two-point function
\begin{equation}
\la
J_{\m}^{\a\b}(x_{1})J_{\n}^{\c\delta}(x_{2})\ra=C_{J}\frac{I_{\m\n}(x_{12})
}{x_{12}^{2(d-1)}}{\textstyle{\frac{1}{2}}}(\delta^{\a\c}\delta^{\b\delta}
-\delta^{\a\delta}\delta^{\b\c})\,,\,\,\,\,\,I_{\m\n}(x)=\delta_{\m\n}
-{\textstyle{
2\,\frac{x_{\m}x_{\n}}{x^{2}}}}\,,\label{eq15}
\end{equation}
is fixed from conformal invariance
\cite{tasos1,tasos2} up 
to a proportionality constant $C_{J}$ which has been proposed as a
possible generalisation of the $k$-theorem in dimensions $d>2$
\cite{tasos2,Cardona}. From (\ref{eq7}), $J^{\a\b}_{\m}(x)$
contributes to the most singular term of $F^{V}_{ijkl}$  in the short
distance expansion
as $x_{12}^{2}$, $x_{34}^{2}\rightarrow 0$. To find
this contribution we need to have an expression for the most singular
term as $x_{12}^{2}\rightarrow 0$ of
$D^{1}(x_{12},\pa_{2})\,J^{\a\b}_{\mu}(x_{2})$. Using the conformally
invariant form of the three-point function\footnote{In general there
are two independent conformal structures in the three-point function
$\la \psi\bpsi J_{\m}\ra$ \cite{Todorov,Nobili} and hence two
independent coupling constants. The Ward identity  gives a
relation between these two constants. Then, our consistency
requirement should presumably determine the remaining coupling constant
{\it{and}} the important parameter $C_{J}$ to leading order in $1/N$.
This calculation  will be
presented in our more detailed forthcoming publication \cite{tasos3}.
However, for   the purposes of 
the present work we only need one of the two above mentioned terms,
namely the one including the most singular contribution as
$x_{12}^{2}\rightarrow 0$, whose coupling constant is denoted by
$g_{\psi\bpsi J}$.}  $\la\psi\bpsi J_{\m}\ra$ we
obtain
\begin{equation}
\psi_{i}^{\a}(x_{1})\bpsi_{j}^{\b}(x_{2})\leadsto\frac{g_{\psi\bpsi
J}}{C_{T}}\frac{1}{(x_{12}^{2})^{a-\m-1}}(\gamma_{\m})
_{ij}J_{\m}^{\a\b}(x_{2})+\cdots\,,\label{eq16}
\end{equation}
where the dots stand for less singular terms. From (\ref{eq16}) we
then obtain for the leading $J_{\m}^{\a\b}(x)$ contribution
\begin{equation}
F^{V}_{ijkl}(x_{1},x_{2},x_{3},x_{4})\leadsto\frac{g_{\psi\bpsi
J}^{2}}{C_{J}}\frac{1}{(x_{12}^{2}x_{34}^{2})^{a-\m}}
(\gamma_{\m})_{ij}\otimes(\gamma_{\n})_{kl}\frac{I_{\m\n}(x_{24})}{
x_{24}^{2(d-1)}}+\cdots\,,\label{eq17}
\end{equation}
where again the dots stand for less singular terms. All that remains
now for completing our calculation is to construct a conformally
invariant graphical expansion for
$\Psi^{\a\b\c\delta}_{ijkl}(x_{1},..,x_{4})$, evaluate the amplitudes
and compare the results with (\ref{eq11}), (\ref{eq14}) and (\ref{eq17}).

\subsection*{The Free Theory}

This is easily done for a theory of free
massless $N$-component fermions where the four-point function
(\ref{eq2})  can be
exactly calculated using Wick's theorem with elementary contraction
(\ref{eq1}). We easily obtain
\begin{eqnarray}
F^{S}_{ijkl}(x_{1},..,x_{4}) & = &
\frac{(\not\!x_{12})_{ij}\otimes(\not\!x_{34})_{kl}}{(x_{12}^{2}x_{34}^{2})
^{a}}-{\textstyle{\frac{1}{N}}}\Bigl[\frac{(\not\!x_{14})_{il}
\otimes(\not\!x_{32})_{kj}}{(x_{14}^{2}x_{23}^{2})^{a}}- \frac{
(\not\! x_{13})_{in}\otimes(\not\!
x_{24})_{mj}}{(x_{14}^{2}x_{23}^{2})^{a}}C_{kn}C^{-1}_{mj}\Bigl]\,
\label{eq180} \\
F^{V}_{ijkl}(x_{1},..,x_{4}) & = & \frac{(\not\!x_{14})_{il}\otimes(\not\!
x_{32})_{kj}}{(x_{14}^{2}x_{23}^{2})^{a}}+\frac{(\not\!x_{13})_{in}\otimes
(\not\!
x_{24})_{mj}}{(x_{14}^{2}x_{23}^{2})^{a}}C_{kn}C^{-1}_{mj}\,.\label{eq18}
\end{eqnarray}
To compare (\ref{eq180}) and (\ref{eq18}) with (\ref{eq11}),
(\ref{eq14}) and
(\ref{eq17}) we use the Fierz identity and the properties of the
charge conjugation matrix and we obtain for the most singular terms in
the limit 
$x_{12}^{2}$, $x_{34}^{2}\rightarrow 0$ 
\begin{eqnarray}
F^{S}_{ijkl}(x_{1},..,x_{4}) & \leadsto &
\frac{(\not\!x_{12})_{ij}\otimes(\not\!x_{34})_{kl}}{(x_{12}^{2}x_{34}^{2})
^{a}}+\frac{1}{(\mbox{Tr}{\bf{1}})N}\frac{\left[(x_{13}\cdot
x_{24})+\left(\frac{v}{u}\right)^{a}(x_{14}\cdot 
x_{23})\right]}{(x_{13}^{2}x_{24}^{2})^{a}}      
\,{\bf{1}}_{ij} \otimes  {\bf{1}}_{kl}+\cdots \,,\non \\
 & = &
\frac{(\not\!x_{12})_{ij}\otimes(\not\!x_{34})_{kl}}{(x_{12}^{2}
x_{34}^{2})^{a}}+ 
\frac{2}{(\mbox{Tr}{\bf{1}})N}\frac{x_{24}^{2}}{(x_{12}^{2}x_{34}^{2})^{a}}
\,v^{a}\,{\bf{1}}_{ij} \otimes  {\bf{1}}_{kl}+\cdots\,, \label{eq19} \\
F^{V}_{ijkl}(x_{1},..,x_{4}) & \leadsto &
\frac{2}{(\mbox{Tr}{\bf{1}})}\frac{x_{24}^{2}}{(x_{12}^{2}
x_{34}^{2})^{a}}(uv)^{
\frac{1}{2}a}\left(\frac{u}{v}\right)^{\frac{1}{2}a 
+ \frac{1}{4}}(\gamma_{\m})_{ij}\otimes (\gamma_{\n})_{kl}
\,I_{\m\n}(x_{24}) +\cdots \,,\non\\
 & = & \frac{2}{(\mbox{Tr}{\bf{1}})}(\gamma_{\m})_{ij}\otimes
(\gamma_{\n})_{kl} \,\frac{I_{\m\n}(x_{24})}{x_{24}^{4a-2}} +\cdots \,,
\label{eq20}
\end{eqnarray}
where we have used 
\begin{equation}
\frac{x_{12}^{2}x_{34}^{2}}{x_{24}^{4}}=(uv)^{\frac{1}{2}}+O(|x_{12}|,
|x_{34}|)\,\,\,,\,\,\,\left(\frac{u}{v}\right)^{k}=1+O(|x_{12}|,|x_{34}|)
\end{equation}
Consistency then of  (\ref{eq20})
with (\ref{eq17}) requires
\begin{equation}
\eta=\m-{\textstyle{\frac{1}{2}}}\,,\,\,\,\,\,\,\,\,\frac{g^{2}_{\psi\bpsi
J}}{C_{J}}=\frac{2}{(\mbox{Tr}{\bf{1}})}\,,\label{eq21}
\end{equation}
while from (\ref{eq180}) and (\ref{eq12}), (\ref{eq14})  we deduce
that 
\begin{equation}
\eta_{o}=2\eta=d-1\,,\,\,\,\,\,\,\,\,g_{\psi\bpsi
O}^{2}=\frac{2}{(\mbox{Tr}{\bf{1}})\,N}\,,\label{eq210}
\end{equation}
in agreement with the usual free field theory results. Note that the
field $O(x)$ in the free field theory OPE (\ref{eq7}) 
can be  identified with the normal product
$:\bpsi^{\a}_{i}(x)\psi^{\a}_{i}(x):/\sqrt{2N}$ as expected.

\subsection*{The Non-Trivial Theory}

As mentioned before, the amplitudes for the non-trivial theory are
constructed  in terms of
skeleton graphs with internal
lines corresponding to the full two-point functions 
$\la\psi\bpsi\ra$ and $\la\tilde{O}\tilde{O}\ra$ while the triple vertices
connecting the lines are   fully amputated three-point functions. 
For example the triple vertex $V_{ij}^{\psi\bpsi
\tilde{O}}(x_{1},x_{2},x_{3})\,\delta^{\a\b}$
between, $\psi^{\a}_{i}(x_{1})$, $\bpsi^{\b}_{j}(x_{2})$ and
$\tilde{O}(x_{3})$ is
obtained on amputating all three legs from the full three-point function  
$\la\psi^{\a}_{i}(x_{1})\bpsi^{\b}_{j}(x_{2})\tilde{O}(x_{3})\ra$: the
latter  is 
analogous to (\ref{eq11}) when
$\eta_{o}\rightarrow\tilde{\eta}_{o}$ and $g_{\psi\bpsi O}\rightarrow
G_{*}$ with $G_{*}$ the coupling. The three-point function may be
graphically represented as shown in Fig.1. 
Then, using the inverse kernels 
\begin{eqnarray}
\left[\la\tilde{O}(x_{1})\tilde{O}(x_{2})\ra\right]^{-1}=\rho(\tilde   
{\eta}_{o})\frac{1}{(x_{12}^{2})^{d-\tilde{\eta}_{o}}} & , &
\rho(x)=\frac{1}{\pi^{d}}\frac{\Gamma(d-x)
\Gamma(x)}{\Gamma(x-\mu)
\Gamma(\mu-x)}\,.\label{eq23} \\
\left[\la\psi^{\a}_{i}(x_{1})\bpsi^{\b}_{j}(x_{2}) \ra \right]^{-1}  =
\mbox{p}(a)
\frac{(\not\!x_{12})_{ij}}{(x_{12}^{2})^{d-a}}\,\delta^{\a\b} & , &
\mbox{p}(a)=
\frac{a-1}{d-a+1}
\,\rho(a-1)
\,,\label{eq22} 
\end{eqnarray}
and the fermionic DEPP formula \cite{DEPP,tasos1,tasos2} 
\begin{eqnarray}
 & & \hspace{-1cm}\int
{\mbox{d}}x\frac{[\gamma\!\cdot\!(x-x_{2})][\gamma\!\cdot\!(x-x_{3})]}
{(x-x_{1})
^{2a_{1}}(x-x_{2})^{2a_{2}}(x-x_{3})^{2a_{3}
}} =
\frac{{\cal{F}}(a_{1};a_{2},a_{3})[\gamma\!\cdot\!(x_{12})][\gamma\!\cdot\!
(x_{13})]}{(x_{23}^{2})^{\mu-a_{1}}
(x_{12}^{2})^{\mu-a_{3}+1}(x_{13}^{2})^{\mu
-a_{2}+1}}\,,\label{eq24} \\
 &&{\cal{F}}(a_{1};a_{2},a_{3}) = 
\pi^{\mu}\frac{\Gamma(\mu-a_{1})\Gamma(\mu-a_{2}+1)
\Gamma(\mu-a_{3}+1)}{
\Gamma(a_{1})\Gamma(a_{2})
\Gamma(a_{3})}\,,\label{eq25}
\end{eqnarray}
which holds only for $a_{1}+a_{2}+a_{3}=d$, 
we obtain
\begin{eqnarray}
& & V_{ij}^{\psi\bpsi
\tilde{O}}(x_{1},x_{2},x_{3}) =
\lambda_{*}\frac{(\not\!x_{13}\not\!x_{32})_{ij}} 
{(x_{12}^{2})^{\mu-a+1}(x_{13}^{2}x_{23}^{2})
^{\mu-\frac{1}{2} \tilde{\eta}_{o}+\frac{1}{2}}}\,,\label{eq26}\\
& & \lambda_{*} =
G_{*}\frac{1}{\pi^{3\mu}}\frac{\Gamma^{3}(\mu-\frac{1}{2}
\tilde{\eta}_{o}
+\frac{1}{2})\,\Gamma^{2}(a)\,\Gamma(\tilde{\eta}
_{o})
\,\Gamma(d-a-\frac{1}{2}+\frac{1}{2}\tilde{\eta}_{o})}{\Gamma^{3}(\frac{1}{2}
\tilde{\eta}_{o}+\frac{1}{2})\,\Gamma^{2}(\mu-a+1)\,\Gamma
(\mu-\tilde{\eta}
_{o})\,\Gamma(a-\m -\frac{1}{2}+\frac{1}{2}\tilde{\eta}_{o})}\,.\label{eq27}
\end{eqnarray}
This amputation can be diagrammatically represented as shown in
Fig.2 where the double directed  line denotes the inverse kernel
(\ref{eq22}), the dotted line the inverse kernel (\ref{eq23}),
the lines ending in a circle are amputated and the dark blob denotes the
full vertex (\ref{eq26}). For simplicity we have factored out the
$O(N)$ indices from Fig.2 
\begin{figure}[t]

\setlength{\unitlength}{0.0080in}%
\begin{picture}(505,160)(100,465)
\thicklines
\put(520,600){\line( 4,-3){ 80}}
\put(600,540){\line(-4,-3){ 80}}
\put(520,480){\line( 0, 1){120}}
\multiput(560,570)(-0.25000,0.50000){21}{\makebox(0.4444,0.6667){\sevrm .}}
\put(560,570){\line(-1, 0){ 10}}
\multiput(560,510)(0.25000,0.50000){21}{\makebox(0.4444,0.6667){\sevrm .}}
\put(560,510){\line( 1, 0){ 10}}
\put(200,535){\makebox(0,0)[lb]{\raisebox{0pt}[0pt][0pt]{\twlrm
$\la\psi_{i}^{\a}(x_{1})\bpsi_{i}^{\b}(x_{2})\tilde{O}(x_{3})\ra\,\,\,\,\,
=$}}} 
\put(500,610){\makebox(0,0)[lb]{\raisebox{0pt}[0pt][0pt]{\twlrm
$x_{1},{\scriptstyle{i}}$ }}}
\put(500,465){\makebox(0,0)[lb]{\raisebox{0pt}[0pt][0pt]{\twlrm
$x_{2},{\scriptstyle{j}}$ }}}
\put(605,520){\makebox(0,0)[lb]{\raisebox{0pt}[0pt][0pt]{\twlrm $x_{3}$}}}
\put(630,535){\makebox(0,0)[lb]{\raisebox{0pt}[0pt][0pt]{\twlrm
$\delta^{\a\b}$
}}}
\end{picture}
\caption{The graphical representation of
$\la\psi^{\a}_{i}(x_{1})\bpsi^{\b}(x_{2})\tilde{O}(x_{3})\ra$.}

\end{figure}

\begin{figure}[t]

\setlength{\unitlength}{0.007in}%
\begin{picture}(340,280)(440,405)
\thicklines
\put(520,600){\line( 4,-3){ 80}}
\put(600,540){\line(-4,-3){ 80}}
\put(520,480){\line( 0, 1){120}}
\multiput(560,570)(-0.25000,0.50000){21}{\makebox(0.4444,0.6667){\rm
.}}
\multiput(560,570)(-0.6000,0.1000){21}{\makebox(0.4444,0.6667){\rm
.}}

\multiput(560,510)(0.25000,0.50000){21}{\makebox(0.4444,0.6667){\rm
.}}
\multiput(560,510)(0.6000,0.10000){21}{\makebox(0.4444,0.6667){\rm
.}}
\put(521,660){\line( 0,-1){ 60}}
\put(518,660){\line( 0,-1){ 60}}
\put(521,480){\line( 0,-1){ 60}}
\put(518,480){\line( 0,-1){ 60}}
\multiput(520,625)(-0.25000,0.50000){21}{\makebox(0.4444,0.6667){\rm .}}
\multiput(520,625)(0.25000,0.50000){21}{\makebox(0.4444,0.6667){\rm .}}
\multiput(520,445)(-0.25000,0.50000){21}{\makebox(0.4444,0.6667){\rm .}}
\multiput(520,445)(0.25000,0.50000){21}{\makebox(0.4444,0.6667){\rm .}}
\multiput(600,540)(5.90909,0.00000){10}{\makebox(0.3,0.3){\rm
.}}
\put(780,600){\circle{10}}
\put(780,600){\line( 1,-1){ 60}}
\put(780,480){\line( 1, 1){ 60}}
\put(780,480){\circle{10}}
\put(840,540){\circle*{10}}
\put(900,540){\circle{10}}
\multiput(840,540)(10.90909,0.00000){6}{\line( 1, 0){  5.455}}
\multiput(810,570)(-0.25000,0.50000){21}{\makebox(0.4444,0.6667){\rm .}}
\multiput(810,570)(-0.50000,0.25000){21}{\makebox(0.4444,0.6667){\rm .}}
\multiput(805,505)(0.25000,0.50000){21}{\makebox(0.4444,0.6667){\rm .}}
\multiput(805,505)(0.50000,0.25000){21}{\makebox(0.4444,0.6667){\rm .}}
\put(700,535){\makebox(0,0)[lb]{\raisebox{0pt}[0pt][0pt]{\twlrm $=$}}}
\put(780,610){\makebox(0,0)[lb]{\raisebox{0pt}[0pt][0pt]{\twlrm
$x_{1},{\scriptstyle{i}}$}}}
\put(895,520){\makebox(0,0)[lb]{\raisebox{0pt}[0pt][0pt]{\twlrm $x_{3}$}}}
\put(780,460){\makebox(0,0)[lb]{\raisebox{0pt}[0pt][0pt]{\twlrm
$x_{2},{\scriptstyle{j}}$}}}
\put(510,670){\makebox(0,0)[lb]{\raisebox{0pt}[0pt][0pt]{\twlrm
$x_{1},{\scriptstyle{i}}$}}}
\put(650,520){\makebox(0,0)[lb]{\raisebox{0pt}[0pt][0pt]{\twlrm $x_{3}$}}}
\put(510,405){\makebox(0,0)[lb]{\raisebox{0pt}[0pt][0pt]{\twlrm
$x_{2},{\scriptstyle{j}}$}}}
\put(950,535){\makebox(0,0)[lb]{\raisebox{0pt}[0pt][0pt]{\twlrm
$=\,\,\,\,\,
V_{ij}^{\psi\bpsi \tilde{O}}(x_{1},x_{2};x_{3})$ 
}}}
\end{picture}
\caption{Amputating the three-point function}
\end{figure}
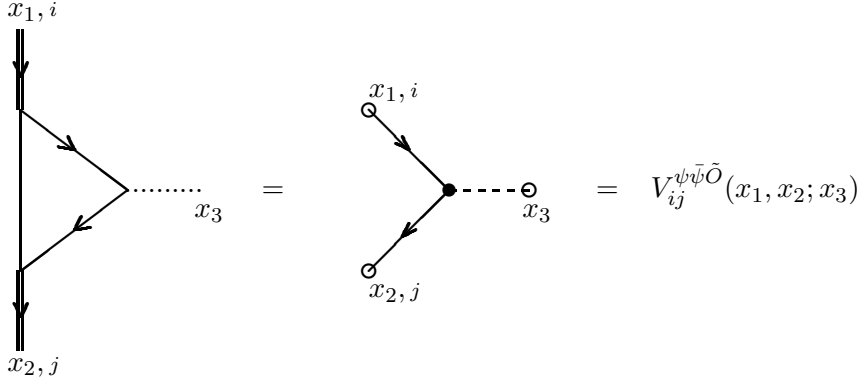
For the illustrative purposes of this letter and according to our
proposed construction the first two terms in the
graphical expansion of
$\Psi_{ij,kl}^{(\tilde{\eta}_{o})}(x_{1},x_{2};x_{3},x_{4})$ in
increasing order according to the number of vertices are as shown in
Fig.3 where the superscript $\tilde{\eta}_{o}$ denotes graphs built
using as internal lines the full two-point function
$\la\tilde{O}\tilde{O}\ra$.
\begin{figure}[t]

\setlength{\unitlength}{0.01400in}%
\begingroup\makeatletter
\def\x#1#2#3#4#5#6#7\relax{\def\x{#1#2#3#4#5#6}}%
\expandafter\x\fmtname xxxxxx\relax \def\y{splain}%
\ifx\x\y   
\gdef\SetFigFont#1#2#3{%
  \ifnum #1<17\tiny\else \ifnum #1<20\small\else
  \ifnum #1<24\normalsize\else \ifnum #1<29\large\else
  \ifnum #1<34\Large\else \ifnum #1<41\LARGE\else
     \huge\fi\fi\fi\fi\fi\fi
  \csname #3\endcsname}%
\else
\gdef\SetFigFont#1#2#3{\begingroup
  \count@#1\relax \ifnum 25<\count@\count@25\fi
  \def\x{\endgroup\@setsize\SetFigFont{#2pt}}%
  \expandafter\x
    \csname \romannumeral\the\count@ pt\expandafter\endcsname
    \csname @\romannumeral\the\count@ pt\endcsname
  \csname #3\endcsname}%
\fi
\endgroup
\begin{picture}(444,120)(-20,570)
\thicklines
\put(165,665){\line( 1, 0){ 60}}
\put(165,615){\line( 1, 0){ 60}}
\put(285,615){\line( 1, 0){ 60}}
\put(285,665){\line( 1, 0){ 60}}
\put(315,665){\circle*{8}}
\put(315,615){\circle*{8}}

\multiput(315,665)(0.00000,-4.00000){13}{\line( 0,-1){  2.000}}
\thinlines
\thicklines
\put(155,670){\makebox(0,0)[lb]{\smash{\SetFigFont{10}{14.4}{rm}$x_{1},
{\scriptstyle{i}}$}}}
\put(225,670){\makebox(0,0)[lb]{\smash{\SetFigFont{10}{14.4}{rm}$x_{2}, 
{\scriptstyle{j}}$}}}
\put(155,605){\makebox(0,0)[lb]{\smash{\SetFigFont{10}{14.4}{rm}$x_{3}
,{\scriptstyle{k}}$}}}
\put(225,605){\makebox(0,0)[lb]{\smash{\SetFigFont{10}{14.4}{rm}$x_{4}
,{\scriptstyle{l}}$}}}
\put(275,605){\makebox(0,0)[lb]{\smash{\SetFigFont{10}{14.4}{rm}$x_{3}
,{\scriptstyle{k}}$}}}
\put(275,670){\makebox(0,0)[lb]{\smash{\SetFigFont{10}{14.4}{rm}$x_{1}
,{\scriptstyle{i}}$}}}
\put(345,670){\makebox(0,0)[lb]{\smash{\SetFigFont{10}{14.4}{rm}$x_{2}
,{\scriptstyle{j}}$}}}
\put(345,605){\makebox(0,0)[lb]{\smash{\SetFigFont{10}{14.4}{rm}$x_{4}
,{\scriptstyle{l}}$}}}
\put(15,637){\makebox(0,0)[lb]{\smash{\SetFigFont{10}{14.4}{rm}$
\Psi^{(\tilde{\eta}_{o})}_{ij,kl}(x_{1},x_{2};x_{3},x_{4})\,\,\,\,=$ }}}
\put(250,637){\makebox(0,0)[lb]{\smash{\SetFigFont{10}{14.4}{rm}$+$}}}

\multiput(200,665)(-0.25000,0.250000){21}{\makebox(0.4,0.6){\rm .}}
\multiput(200,665)(-0.25000,-0.25000){21}{\makebox(0.4,0.6){\rm .}}
\multiput(200,615)(-0.25000,0.25000){21}{\makebox(0.4,0.6){\rm .}}
\multiput(200,615)(-0.25000,-0.25000){21}{\makebox(0.4,0.6){\rm .}}

\multiput(305,665)(-0.25000,0.250000){21}{\makebox(0.4,0.6){\rm .}}
\multiput(305,665)(-0.25000,-0.250000){21}{\makebox(0.4,0.6){\rm .}}
\multiput(332,665)(-0.25000,0.250000){21}{\makebox(0.4,0.6){\rm .}}
\multiput(332,665)(-0.25000,-0.250000){21}{\makebox(0.4,0.6){\rm .}}
\multiput(305,615)(-0.25000,0.250000){21}{\makebox(0.4,0.6){\rm .}}
\multiput(305,615)(-0.25000,-0.250000){21}{\makebox(0.4,0.6){\rm .}}
\multiput(332,615)(-0.25000,0.250000){21}{\makebox(0.4,0.6){\rm .}}
\multiput(332,615)(-0.25000,-0.250000){21}{\makebox(0.4,0.6){\rm .}}

\put(370,637){\makebox(0,0)[lb]{\smash{\SetFigFont{10}{14.4}{rm} $ +
\cdots$}}}
\put(190,685){\makebox(0,0)[lb]{\smash{\SetFigFont{10}{14.4}{rm}${
\cal{G}}^{0}_{ij,kl}$}}}
\put(310,685){\makebox(0,0)[lb]{\smash{\SetFigFont{10}{14.4}{rm}${
\cal{G}}^{1,(\tilde{\eta}_{o})}_{ij,kl}$}}}

\end{picture}

\caption{The Skeleton Graph Expansion for
$\Psi^{(\tilde{\eta}_{o})}_{ij,kl}(x_{1},x_{2};x_{3},x_{4})$}\label{fg4}

\end{figure}
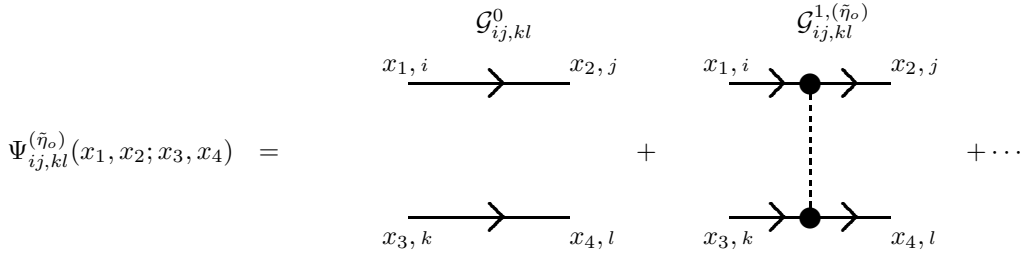
After some algebra whose details follow closely the calculations in
\cite{tasos1,tasos2} we obtain
\begin{equation}
{\cal{G}}^{1,(\tilde{\eta}_{o})}_{ij,kl}(x_{1},x_{2};x_{3},x_{4})= 
G_{*}^{2}\frac{x_{24}^{2}}{(x_
{12}^{2}x_{34}^{2})^{a}}\left[{\cal{A}}^{(\tilde{\eta}_{o})}_{ij,kl}
(x_{1},..,x_{4})+C_{F}
(d-\tilde{\eta}_{o})\,{\cal{A}}^{(d-\tilde{\eta}_{o})}_{ij,kl} (x_{1}
,..,x_{4}
)\right]\,, \label{eq28}
\end{equation}
with ${\cal{A}}^{(\tilde{\eta}_{o})}_{ij,kl}(u,v)$ as in (\ref{eq12}) and 
\begin{equation}
C_{F}(d-\tilde{\eta}_{o})=C_{F}^{-1}(\tilde{\eta}_{o})= 
\frac{\Gamma(\tilde{\eta}_{o})     
\Gamma(\tilde{\eta}_{o}-\mu)\Gamma^{4}(\mu-\frac{1}{2}\tilde{\eta}_{o}    
+\frac{1}{2})}{\Gamma(d-\tilde{\eta}_{o})\Gamma(\mu-\tilde{\eta}_{o}) 
\Gamma^{4}(\frac{1}{2}\tilde{\eta}_{o}+\frac{1}{2})}\,.\label{eq29}
\end{equation}
Note the similarity of (\ref{eq28}) with the corresponding result for
the one-particle exchange graphs in \cite {tasos1}. Basically, as
pointed out in
that reference, the amplitude for ${\cal{G}}^{1,(\tilde{\eta}_{o})}$
involves contributions from both $\tilde{O}(x)$ and its {\it{shadow
field}} \footnote{For the notion  of {\it{shadow fields}} which
correspond to the {\it{shadow symmetry}} property of the conformal
group in $d>2$ see \cite{Parisi} and references therein.}
$\tilde{O}_{S}(x)$  with dimension $d-\tilde{\eta}_{o}$.

From (\ref{eq4}), (\ref{eq18}) and (\ref{eq28}) we obtain 
\begin{eqnarray}
F^{S}_{ijkl}(x_{1},..,x_{4}) & = &
\frac{(\not\!x_{12})_{ij}\otimes(\not\!x_{34})_{kl}}{(x_{12}^{2}x_{34}^{2})
^{a}}-{\textstyle{\frac{1}{N}}}\Bigl[\frac{(\not\!x_{14})_{il}\otimes(\not\!
x_{32})_{kj}}{(x_{14}^{2}x_{23}^{2})^{a}}-\frac{(\not\!x_{13})_{in}\otimes
(\not\!
x_{24})_{mj}}{(x_{14}^{2}x_{23}^{2})^{a}}C_{kn}C^{-1}_{mj}\Bigl]\,\non
\\
 & + & G_{*}^{2}\frac{x_{24}^{2}}{(x_
{12}^{2}x_{34}^{2})^{a}}\left[{\cal{A}}^{(\tilde{\eta}_{o})}_{ij,kl}
(x_{1},..,x_{4})+C_{F}
(d-\tilde{\eta}_{o}){\cal{A}}^{(d-\tilde{\eta}_{o})}_{ij,kl}(x_{1},..,x_{4})
\right]
+\cdots \,.    
 \label{eq30}
\end{eqnarray}
Equation (\ref{eq30}) must agree with
(\ref{eq11}) and one immediately observes a remarkable correspondence:
(\ref{eq11}) and (\ref{eq30}) contain all-order contributions in the
expansion as $x_{12}^{2}$, $x_{34}^{2}\rightarrow 0$ with very similar
closed analytic form. We refer to the second term on the r.h.s. of
(\ref{eq11}) and the first term in the second row  of (\ref{eq30}).
Therefore  it seems natural to us to assume that these two terms are in
fact equal, and this can be achieved if we identify $g_{\psi\bpsi
O}\equiv G_{*}$ and $\eta_{o}\equiv\tilde{\eta}_{o}$. But then, from
the results in  the free field theory case, the bracketed term in the first
row of (\ref{eq30}) when expanded as $x_{12}^{2}$,
$x_{34}^{2}\rightarrow 0$ must cancel the leading term of
${\cal{A}}^{(d-\tilde{\eta}_{o})}_{ij,kl}$ in the same limit. It is then
easy to find  that this consistency condition requires  
\begin{equation}
G_{*}^{2}\equiv g_{\psi\bpsi
O}^{2}=-\frac{2}{N}\,C_{F}(\tilde{\eta}_{o})\,\,\,\,\,\,
\mbox{and}\,\,\,\,\,\,2\eta=d-\tilde{\eta}_{o}\equiv d-\eta_{o}\,.  
\label{eq31}
\end{equation}
Equation (\ref{eq31}) is the starting point for a self consistent
solution of the theory. Firstly, since $G_{*}^{2}= O(1/N)$, a
consistent $1/N$ expansion can be constructed:
the order in $1/N$ of the graphs simply corresponds to the number of
triple vertices and at the same time we  expand the couplings and the
dimensions 
of the fields in a canonical part, (which equals their corresponding
free field theory values),  and $1/N$ corrections. Secondly, the
value for $G_{*}^{2}$ in (\ref{eq31}) serves as a ``seed'' in
explicit calculations of the anomalous dimensions and other
interesting quantities of the theory \cite{tasos1,tasos2}. Note that
for $0<\tilde{\eta}_{o}<d$ and $2<d<4$ then  $G_{*}^{2}>0$ which
ensures the  reality of the coupling in 
the graphically constructed theory at least to the order considered
here.  

Clearly,
many interesting calculations can be done in the fermionic
model in hand both to check our approach and also to obtain new interesting
results \cite{tasos3}. Here we only present  evidence that our approach is
consistent with the standard $1/N$ expansion results
\cite{Gracey,Semenoff} by evaluating the $1/N$ correction to the
anomalous dimension $\eta$ of the fermion field. This is most easily
done evaluating  $F^{V}_{ijkl}$ to $O(1/N)$ using our graphical
construction. This involves the evaluation of the ``crossed''
one-particle exchange graphs 
${\cal{G}}^{1,({\tilde{\eta}_{o}})}_{il,kj}(x_{1};x_{4},x_{3};x_{2})$
and  ${\cal{G}}^{1,({\tilde{\eta}_{o}})}_{in,ml}(x_{1};x_{3},x_{2};x_{2})
\,C_{kn}\,C^{-1}_{mj}$ to leading order in $1/N$, i.e. using for the
various parameters their free field theory values (\ref{eq21}),
(\ref{eq210}). This calculation is done using techniques developed in
\cite{DEPP,tasos1,tasos3} and the results are given in the
Appendix. Then we set 
\begin{equation}
\eta = \m
-{\textstyle{\frac{1}{2}}}+{\textstyle{\frac{1}{N}}}\eta_{1}\,,\label{eq32}
\end{equation}
and expand (\ref{eq20}) in $1/N$ when after the inclusion of the
results in the Appendix we obtain \footnote{Here we only present the
logarithmic $O(1/N)$ terms which are  the relevant ones 
for the calculation of the anomalous dimension. The non-logarithmic
$O(1/N)$ terms are needed in the calculation of the couplings
\cite{tasos1} and $C_{J}$ \cite{tasos1,tasos3}.}   
\begin{eqnarray}
F^{V}_{ij,kl}(x_{1},..,x_{4}) & = & \frac{2}{(\mbox{Tr}{\bf{1}})}
\frac{x_{24}^{2}}{(x_{12}^{2}x_{34}^{2})^{a}}\,(uv)^{\frac{1}{2}\m}\left[
1+{\textstyle{\frac{1}{2N}}}\,\eta_{1}\,\mbox{ln}(uv)\right]\,
(\gamma_{\m})_{ij}\otimes(\gamma_{\n})_{kl}\,I_{\m\n}(x_{24})\,,\non \\
& &{}\hspace{-1.5cm}-\frac{G_{*}^{2}}{(\mbox{Tr}{\bf{1}})}\frac{(\m-1)}
{\Gamma(\mu-1)}\,
a_{00}\,\frac{x_{24}^{2}}{(x_{12}^{2}x_{34}^{2})^{a}}\,
(uv)^{\frac{1}{2}\m}\, \mbox{ln}(uv)\,(\gamma_{\m})_{ij}
\otimes(\gamma_{\n})_{kl}\,I_{\m\n}(x_{24})+\cdots\,,\label{eq33}
\end{eqnarray}  
with $a_{nm}$ given in equation (\ref{ap7}) of the Appendix and
\begin{equation}
G_{*}^{2}=-\frac{2}{N}\,C_{F}(1)=\frac{2}{N}\frac{\Gamma(2\m-1)}{
\Gamma(2-\m)\Gamma^{3}(\m)}\,.
\end{equation}
 Consistency of
the logarithmic $O(1/N)$ terms in (\ref{eq33}) and (\ref{eq17}) then yields
\begin{equation}
\eta_{1}=\frac{2}{(\mbox{Tr}{\bf{1}})}\frac{
\Gamma(2\m-1)}{\Gamma^{2}(\m-1)\Gamma(2-\m)
\Gamma(\m+1)}\,, \label{eq34}
\end{equation}
which is in agreement with the value of 
the critical exponent for the fermions in the Gross-Neveu and
four-fermion models in 
$2<d<4$ \cite{Gracey,Semenoff,Stepanenko}. This shows that the our
fermionic model is in  the same universality class with  the above
two models in $2<d<4$ as one expects.

\section*{Summary and Outlook}

We have outlined our approach to CFT in $d>2$ and applied it to a
$O(N)$ invariant fermionic model in $2<d<4$. Assuming the existence of
a ``low-lying 
scalar field'' in the OPE $\psi\bpsi$ and constructing a graphical
expansion for the four-point function $\langle\psi\bpsi\psi\bpsi\ra$
we obtained various consistency relations. These determine the leading
order value in $1/N$ of the critical coupling $G_{*}$ in the theory.
Using this value, we obtained the $1/N$ correction to the anomalous
dimension of the fermion field. 

The calculations in the present work resemble in many ways the
calculations in \cite{tasos1} where the $O(N)$ invariant vector
model was studied. Consequently, a unified picture for CFT's in
$d>2$ emerges, while agreement of (\ref{eq34}) with the customary
$1/N$  expansion
results is  positive evidence for the validity of such a picture.
Our approach emphasises the role played by the
{\it{shadow symmetry}} of the conformal group in  $d>2$, for 
understanding CFT's  in more than two dimensions. Note
that although our  picture for CFT's gives trivial results for
$d=4$,  (e.g. $\eta_{1} = 0$ and $G_{*}^{2}=2/(N\,\mbox{Tr}{\bf{1}})$
in $d=4$),  a
similar picture has
been recently proposed 
for non-trivial four dimensional supersymmetric CFT's
\cite{Grisaru}. A more detailed presentation of the calculations in
the present letter and some new results concerning important
quantities of the $O(N)$ invariant fermionic model
will be presented in a forthcoming publication \cite{tasos3}.

\subsection*{Acknowledgements}
\small
I am indebted to Hugh Osborn for his interest in my work and many  useful
suggestions.
\normalsize

\appendix

\section*{Appendix}

We give here the analytic expressions for the ``crossed'' one-particle
exchange graphs
${\cal{G}}^{1,(\tilde{\eta}_{o})}_{il,kj}$ and
${\cal{G}}^{1,(\tilde{\eta}_{o})}_{in,ml}\,C_{kn}\,C^{-1}_{mj}$ which
are obtained using Symanzik's method \cite{DEPP,tasos3} for
conformal integration.
\begin{eqnarray}
 & & 
{\cal{G}}^{1,(\tilde{\eta}_{o})}_{il,kj}(x_{1};x_{4},x_{3};x_{2}) =
g_{*}^{2}\frac{\Gamma(\tilde{\eta}_{o})}{\Gamma(\m-\tilde{\eta}_{o})
\Gamma^{4}(\frac{1}{2}\tilde{\eta}_{o}+\frac{1}{2})}
\frac{1}{(x_{14}^{2}x_{23}^{2})^{a}}\,,\non \\
 & &\hspace{0.5cm}
\times\Biggl[\frac{x_{24}^{2}}{x_{14}^{2}x_{23}^{2}}(\not
\!x_{13}\not\!x_{43})_{il}\otimes(\not\!x_{13}\not\!x_{12})_{kl}\,
\,{\cal{I}}_{1}(x_{1} ;
{\textstyle{\frac{1}{2}}}\tilde{\eta}_{o}+1,x_{2} ; \m-{\textstyle{ 
\frac{1}{2}}}\tilde{\eta}_{o},x_{3} ;
\m-{\textstyle{\frac{1}{2}}}\tilde{\eta}_{o}+1, x_{4} ; 
{\textstyle{\frac{1}{2}}}\tilde{\eta}_{o})\,,\non \\
 & &\hspace{0.5cm}
+\frac{x_{34}^{2}}{(x_{14}^{2}x_{23}^{2})^{\frac{1}{2}}}
(\not\!x_{12}\not\!x_{42})_{il}\otimes(\not\!x_{13}\not\!x_{12})_{kj}\,
{\cal{I}}_{2}(x_{1} ;
{\textstyle{\frac{1}{2}}}\tilde{\eta}_{o}+1,x_{2} ; \m-{\textstyle{ 
\frac{1}{2}}}\tilde{\eta}_{o}+1,x_{3} ;
\m-{\textstyle{\frac{1}{2}}}\tilde{\eta}_{o}, x_{4} ; 
{\textstyle{\frac{1}{2}}}\tilde{\eta}_{o})\,,\non \\
 & & \hspace{0.5cm}
{}+\frac{1}{x_{34}^{2}}(\not\!x_{13}\not\!x_{43})_{il} \otimes
(\not\!x_{34}\not \!x_{24})_{kj}
{\cal{I}}_{2}(x_{1} ; {\textstyle{\frac{1}{2}}}\tilde{\eta}_{o}, x_{2}
; \m-{\textstyle{
\frac{1}{2}}}\tilde{\eta}_{o},x_{3} ; \m-{\textstyle{\frac{1}{2}}}
\tilde{\eta}_{o}+1, x_{4} ; 
{\textstyle{\frac{1}{2}}}\tilde{\eta}_{o}+1)\,,\non \\
& & \hspace{0.5cm} + \frac{1}{x_{24}^{2}} \,
(\not\!x_{12}\not\!x_{42})_{il} \otimes (\not\!x_{34}\not\!x_{24})_{kj}\,
{\cal{I}}_{2}(x_{1} ; {\textstyle{\frac{1}{2}}}\tilde{\eta}_{o}, x_{2}
; \m-{\textstyle{
\frac{1}{2}}}\tilde{\eta}_{o}+1,x_{3} ;
\m-{\textstyle{\frac{1}{2}}}\tilde{\eta}_{o}, x_{4} ; 
{\textstyle{\frac{1}{2}}}\tilde{\eta}_{o}+1)\,,\non \\
 & & \hspace{0.5cm}  {}-\Bigl({\textstyle{\frac{1}{2}}}(\not\!x_{14}
\gamma_{\m})_{il}\otimes(\gamma_{\m}\not\!x_{32})_{kj}+ {\bf{1}}_{il}
\otimes(\not\!x_{13}\not\!x_{32})_{kj}+(\not\!x_{13}\not\!x_{43})_{il}
\otimes{\bf{1}}_{kj}\Bigl)\non \\
 & & \hspace{1cm}{}\times\, {\cal{I}}_{0}(x_{1} ; {\textstyle{
\frac{1}{2}}}\tilde{\eta}_{o},x_{2} ; \m-{\textstyle{ 
\frac{1}{2}}}\tilde{\eta}_{o},x_{3} ; \m - 
{\textstyle{\frac{1}{2}}}\tilde{\eta}_{o}, x_{4} ; 
{\textstyle{\frac{1}{2}}}\tilde{\eta}_{o})\Biggl],\label{ap1}
\end{eqnarray}
and 
\begin{eqnarray}
& &
{\cal{G}}^{1,(\tilde{\eta}_{o})}_{in,ml}(x_{1};x_{3},x_{2}; x_{4})
\,C_{kn}\, C^{-1}_{mj}  = 
 g_{*}^{2}\frac{\Gamma(\tilde{\eta}_{o})}{\Gamma(\m-\tilde{\eta}_{o})
\Gamma^{4}(\frac{1}{2}\tilde{\eta}_{o}+\frac{1}{2})}
\frac{1}{(x_{13}^{2}x_{24}^{2})^{a}}\,,\non \\
 & &
\hspace{0.5cm}{}\times 
\Biggl[\frac{x_{24}^{2}}{x_{13}^{2}x_{24}^{2}}(\not 
\!x_{14}\not\!x_{34})_{in}\otimes(\not\!x_{12}\not\!x_{14})_{ml}\,
\,{\cal{I}}_{2}(x_{1} ; {\textstyle{\frac{1}{2}}}\tilde{\eta}_{o}+1,
x_{2} ; \m-{\textstyle{
\frac{1}{2}}}\tilde{\eta}_{o},x_{3} ;
{\textstyle{\frac{1}{2}}}\tilde{\eta}_{o}, x_{4} ; 
\m - {\textstyle{\frac{1}{2}}}\tilde{\eta}_{o}+1)\,,\non \\
 & &
\hspace{0.5cm}{}+\frac{x_{34}^{2}}{(x_{13}^{2}x_{24}^{2})^{\frac{1}{2}}}
(\not\!x_{12}\not\!x_{32})_{in}\otimes(\not\!x_{12}\not\!x_{14})_{ml}\,
{\cal{I}}_{2}(x_{2} ; {\textstyle{\frac{1}{2}}}\tilde{\eta}_{o}+1,
x_{2} ; \m-{\textstyle{
\frac{1}{2}}}\tilde{\eta}_{o}+1, x_{3} ;
{\textstyle{\frac{1}{2}}}\tilde{\eta}_{o}, x_{4} ; 
\m - {\textstyle{\frac{1}{2}}}\tilde{\eta}_{o})\,,\non \\
 & & \hspace{0.5cm}
{}+\frac{1}{x_{23}^{2}}(\not\!x_{12}\not\!x_{32})_{in} \otimes
(\not\!x_{32}\not \!x_{34})_{ml}\,
{\cal{I}}_{2}(x_{1} ; {\textstyle{\frac{1}{2}}}\tilde{\eta}_{o}, x_{2}
; \m-{\textstyle{
\frac{1}{2}}}\tilde{\eta}_{o}+1, x_{3} ; {\textstyle{\frac{1}{2}}}
\tilde{\eta}_{o}+1,  x_{4} ; 
\m - {\textstyle{\frac{1}{2}}}\tilde{\eta}_{o})\,,\non \\
& & \hspace{0.5cm} {}+ \frac{1}{x_{34}^{2}} \,
(\not\!x_{14}\not\!x_{34})_{in} \otimes (\not\!x_{32}\not\!x_{34})_{ml}\,
{\cal{I}}_{2}(x_{1} ; {\textstyle{\frac{1}{2}}}\tilde{\eta}_{o}, x_{2}
; \m-{\textstyle{
\frac{1}{2}}}\tilde{\eta}_{o}, x_{3} ;
{\textstyle{\frac{1}{2}}}\tilde{\eta}_{o}+1, x_{4} ; 
\m - {\textstyle{\frac{1}{2}}}\tilde{\eta}_{o}+1)\,,\non \\
 & & \hspace{0.5cm}  {}-\Bigl({\textstyle{\frac{1}{2}}}(\not\!x_{13}
\gamma_{\m})_{in}\otimes(\gamma_{\m}\not\!x_{24})_{ml}+ {\bf{1}}_{in}
\otimes(\not\!x_{32}\not\!x_{24})_{ml}+(\not\!x_{12}\not\!x_{32})_{in}
\otimes{\bf{1}}_{mj}\Bigl) \non \\
 & & \hspace{1cm}\times \,{\cal{I}}_{0}(x_{1} ; {\textstyle{
\frac{1}{2}}}\tilde{\eta}_{o}, x_{2} ; \m - {\textstyle{ 
\frac{1}{2}}}\tilde{\eta}_{o}, x_{3} ; 
{\textstyle{\frac{1}{2}}}\tilde{\eta}_{o},  x_{4} ; 
\m-{\textstyle{\frac{1}{2}}}\tilde{\eta}_{o})\Biggl ]C_{kn}\,  
C^{-1}_{mj} \, ,\label{ap2}
\end{eqnarray}
where the conformal integrals are
\begin{eqnarray}
& & \hspace{-1.5cm}
{\cal{I}}_{0}(x_{1} ; a_{1}, x_{2} ; a_{2}, x_{3} ; a_{3}, x_{4} ;
a_{4})  \int_{0}^{\infty} 
d\lambda_{1}..
d\lambda_{4}\prod_{i=1}^{4}[\lambda_{i}^{a_{i}-\frac{1}{2}}] \,
(S_{\lambda})^{-\m-1}\,\mbox{exp}\Bigl[ -\frac{1}{S_{\lambda}}
\sum_{i\neq j=1}^{4}(\lambda_{i}\lambda_{j}
x_{ij}^{2})\Bigl],\label{ap3}\\
 & & \hspace{-1.5cm}
{\cal{I}}_{2}(x_{1} ; b_{1}, x_{2} ; b_{2},x_{3} ; b_{3}, x_{4} ;
b_{4}) = \int_{0}^{\infty} d\lambda_{1}..
d\lambda_{4}\prod_{i=1}^{4}[\lambda_{i}^{b_{i}-\frac{1}{2}}] \,
(S_{\lambda})^{-\m-2}\,\mbox{exp}\Bigl[ -\frac{1}{S_{\lambda}}
\sum_{i\neq j=1}^{4}(\lambda_{i}\lambda_{j}
x_{ij}^{2})\Bigl],\label{ap4}\\
 & & \hspace{2cm} S_{\lambda} = \sum_{i=1}^{4}\lambda_{i}\,.
\end{eqnarray}
The integrals (\ref{ap3}) and (\ref{ap4}) are conformally invariant
only if $\sum_{i=1}^{4}a_{i}=d$ and $\sum_{i=1}^{4}b_{i}=d+2$
respectively. Here we only present the result for
${\cal{I}}_{0}$ which is the only integral  involved in the  
calculation of the 
leading terms in (\ref{ap1}), (\ref{ap2}) as $x_{12}^{2}$,
$x_{34}^{2}\rightarrow 0$.
\begin{eqnarray}
{\cal{I}}_{0}(x_{1} ; a_{1}, x_{2} ; a_{2}, x_{3} ; a_{3}, x_{4} ;
a_{4}) & = &  \frac{1}{(x_{23}^{2})^{\m-a_{4}+\frac{1}{2}}
(x_{14}^{2})^{a_{1} +\frac{1}{2}} (x_{34}^{2})^{a_{3}+a_{4}-\m}
(x_{23}^{2})^{a_{2}+a_{3}-\m}} \non \\
 & & {}\hspace{-5cm} \times\,\Biggl[ \sum_{n=0}^{\infty} \frac{v^{n}}{n!}
\frac{\Gamma(a_{3} 
+ a_{4}-\m)\Gamma(\m-a_{3}+\frac{1}{2} +n)\Gamma(a_{2} +\frac{1}{2}+n)
\Gamma(\m -a_{4}+\frac{1}{2}+n)
\Gamma(a_{1}+\frac{1}{2}+n)}{(a_{1}+ a_{2}+1-\m)_{n}\Gamma(a_{1} +
a_{2} +1 +2n)} \non \\
 & & {}\hspace{-4.5cm}\times
{}_{2}F_{1}(\m-a_{4}+{\textstyle{\frac{1}{2}}}+n,a_{1} + 
{\textstyle{\frac{1}{2}}} +n ; a_{1}+ a_{2} + 1 + 2n;
1-{\textstyle{\frac{v}{u}}})\non \\
 & & {}\hspace{-4.5cm} +\,v^{a_{3} +a_{4} -\m}\, \sum_{n=0}^{\infty}
\frac{v^{n}}{n!}\frac{\Gamma(\m - a_{3}  
- a_{4})\Gamma(a_{4}+\frac{1}{2} +n)\Gamma(\m - a_{1} +\frac{1}{2}+n)
\Gamma(a_{3}+\frac{1}{2}+n)}{(a_{3}+ a_{4}+1-\m)_{n}\Gamma(a_{3} +
a_{4} +1 +2n)} \non \\
 & & {}\hspace{-4.5cm}\times\Gamma(\m -
a_{2}+{\textstyle{\frac{1}{2}}}+n)\,
{}_{2}F_{1}(a_{3}+{\textstyle{\frac{1}{2}}}+n,\m -a_{2} + 
{\textstyle{\frac{1}{2}}} +n ; a_{3}+ a_{4} + 1+2n;
1-{\textstyle{\frac{v}{u}}}) \Biggl]\,. \label{ap5}
\end{eqnarray}
For (\ref{ap1}) and (\ref{ap2}) we need  
(\ref{ap5}) when  $a_{1}+a_{2}=a_{3}+a_{4}=\m$  which yields
\begin{eqnarray}
 & & \hspace{-3cm}
{\cal{I}}_{0}(x_{1};{\textstyle{\frac{1}{2}}}\tilde{\eta}_{o} , x_{2} 
; \m - {\textstyle{\frac{1}{2}}}\tilde{\eta}_{o} , x_{3} ; \m -
{\textstyle{\frac{1}{2}}}\tilde{\eta}_{o} , x_{4} ;
{\textstyle{\frac{1}{2}}}\tilde{\eta}_{o}) \non \\
 & & {}=\frac{1}{x_{23}^{\m-\tilde{\eta}_{o}}}
\frac{1}{(x_{14}^{2}x_{23}^{2})^{\frac{1}{2}\tilde{\eta}_{o}+\frac{1}{2}}}
\sum_{n,m=0}^{\infty}
\frac{v^{n}(1-\frac{v}{u})^{m}}{n!m!}\,a_{nm}[-\mbox{ln}v +
b_{nm}]\,,\label{ap6} 
\end{eqnarray}
with
\begin{eqnarray}
a_{nm} & = &  \frac{\Gamma(\frac{1}{2}\tilde{\eta}_{o}+\frac{1}{2}+n)
\Gamma(\m -\frac{1}{2}\tilde{\eta}_{o}+\frac{1}{2}+n)
\Gamma(\m-\frac{1}{2}\tilde{\eta}_{o} + n + m)
\Gamma(\frac{1}{2}\tilde{\eta}_{o} +\frac{1}{2}+n+m)}{\Gamma(1+n)
\Gamma(\m+1 + 2n +m)} \,,\label{ap7} \\
b_{nm} & = & 2\Psi(1+n)+2\Psi(\m+1+2n+m)
-\Psi({\textstyle{\frac{1}{2}}}\tilde{\eta}_{o}
+{\textstyle{\frac{1}{2}}}+n)-\Psi(\m
-{\textstyle{\frac{1}{2}}}\tilde{\eta}_{o}+{\textstyle{\frac{1}{2}}}
+n) \non \\
 & & {}- \Psi(\m
-{\textstyle{\frac{1}{2}}}\tilde{\eta}_{o}+{\textstyle{\frac{1}{2}}}
+n +m ) -
\Psi({\textstyle{\frac{1}{2}}}\tilde{\eta}_{o}+{\textstyle{\frac{1}{2}}} 
+n +m)\,,\label{ap8}
\end{eqnarray}
where $\Psi(x)=\Gamma'(x)/\Gamma(x)$. ${\cal{I}}_{0}(x_{1};{
\textstyle{\frac{1}{2}}}\tilde{\eta}_{o} ,  
x_{2} ; \m - {\textstyle{\frac{1}{2}}}\tilde{\eta}_{o} , x_{3} ;  
{\textstyle{\frac{1}{2}}}\tilde{\eta}_{o} , x_{4} ;\mu -
{\textstyle{\frac{1}{2}}}\tilde{\eta}_{o})$ is obtained from
(\ref{ap6}) setting $x_{3}\leftrightarrow x_{4}$.

\end{document}